# Does presence of social media plugins in a journal website result in higher social media attention of its research publications?

Mousumi Karmakar[a], Sumit Kumar Banshal[b] & Vivek Kumar Singh[a,1]

[a] Department of Computer Science, Banaras Hindu University, Varanasi-221005, India
[b] Department of Computer Science, South Asian University, New Delhi-110021, India

**Abstract:** Social media platforms have now emerged as an important medium for wider dissemination of research articles; with authors, readers and publishers creating different kinds of social media activity about the article. Some research studies have even shown that articles that get more social media attention may get higher visibility and citations. These factors are now persuading journal publishers to integrate social media plugins in their webpages to facilitate sharing and dissemination of articles in social media platforms. Many past studies have analyzed several factors (like journal impact factor, open access, collaboration etc.) that may impact social media attention of scholarly articles. However, there are no studies to analyze whether the presence of social media plugin in a journal could result in higher social media attention of articles published in the journal. This paper aims to bridge this gap in knowledge by analyzing a sufficiently large-sized sample of 99,749 articles from 100 different journals. Results obtained show that journals that have social media plugins integrated in their webpages get significantly higher social media mentions and shares for their articles as compared to journals that do not provide such plugins. Authors and readers visiting journal webpages appear to be a major contributor to social media activity around articles published in such journals. The results suggest that publishing houses should actively provide social media plugin integration in their journal webpages to increase social media visibility (altmetric impact) of their articles.

## Introduction

Scholarly articles are now disseminated and shared on different social media platforms such as Twitter, Facebook, LinkedIn as well as academic social networks like Academia, Mendeley and ResearchGate. These social and academic social networks provide wide-range of facilities which can be useful for academics (Gruzd & Goertzen, 2013), with some studies (Williams & Woodacre, 2016) pointing out that these social and academic networks are informative and relevant for quantitative characterization of research assessments. The activity of measuring the interaction of social media with scholarly information processing is now a well-established area, known as altmetrics (Priem, 2014; Priem et al., 2010; Priem & Hemminger, 2010). Over the last few years, altmetrics has attracted a lot of attention from different quarters, including research performance assessors, authors and journal publishers. Owing to the popularity of altmetrics and increased social media attention to scholarly articles, many journals have now integrated social media plugins in their web pages to facilitate sharing and dissemination of articles in social media platforms.

---
[1] Corresponding Author. Email: vivek@bhu.ac.in

Several previous studies on altmetrics have tried to analyze the correlation between altmetrics and citations, many providing the evidence that altmetrics are positively correlated to citations. These studies used data from a variety of social platforms and/ or academic social networks and found a varying degree of correlations between altmetrics and citations, ranging from weak positive to strong positive. Data samples drawn from different countries/ regions, journals and subjects were analyzed in different studies and the overall understanding is that there exist positive correlations between altmetrics and citations. Due to the increasing social media activity around scholarly articles and the positive evidence of the correlation between altmetrics and citations, many studies have also tried to analyze what factors or article characteristics may result in higher social media attention of scholarly articles. The factors analyzed in previous studies include journal impact factor, international collaboration, open access etc. Studies have suggested that international collaboration and open access are both found to be positively associated with higher altmetric impact of articles, whereas no conclusive evidence could be found about journal impact factor. One important factor that has, however, remained totally unexplored is the presence or absence of social media plugin in journal webpages.

Journals are now integrating different social media plugins directly in their webpages to facilitate social sharing and dissemination of information about articles. Many journal publishing houses now provide for integration of well-known social platforms like Facebook, Mendeley, LinkedIn, Twitter etc. in their journal webpages. These plugins are usually integrated in journal websites in the form of icons that contain a link to the respective social media platform. By clicking the link, a user may easily tweet, like, share or comment on an article in the journal. As a practice, journals which have these plugins integrated also provide different kinds of counts and scores about social shares/ mentions of articles. It would therefore, be interesting to analyze what could be the impact of these plugins on social media visibility of articles. In the absence of any studies on this aspect, it is not clear whether the presence of a social plugin in journal webpages actually results in higher social media activity around articles in that journal. This paper aims to explore this question by analyzing a sufficiently large-sized sample of articles published in journals of both kinds: those with social media plugins and those without the plugins. The journals used for the analysis are selected in a manner that all major disciplines and publishing houses are represented. The coverage and average mention values in different social platforms for the articles in these two sets of journals are analyzed to observe the possible impact of the presence/ absence of social media plugins in journal webpages.

**Research Question**

The main research question being explored is:

*RQ: Does presence of a social media plugin in a journal result in higher social media attention to articles published in that journal?*

Social media data for a large-sized data sample comprising of 99,749 articles drawn from 100 journals, 50 with social media plugins integrated and 50 without the plugins integrated, is analyzed to find an answer to this research question.

**Related Work**

Over the last 6-7 years, different kinds of research studies have been carried out in altmetrics ranging from country-specific studies (Banshal et al., 2018, 2019a; Hammarfelt, 2014) to discipline-specific studies (Bar-ilan, 2014; Chen et al., 2015; Htoo & Na, 2017; Sotudeh et al., 2015; Vogl et al., 2018). Some previous studies also tried to analyze the connection between discipline and social media attention levels of scholarly articles (Banshal et al., 2019b; Costas et al., 2015b, 2015a; Ortega, 2015; Thelwall & Kousha, 2017) and found that disciplinary variations exist in social media coverage of scholarly articles in different platforms.

One of the most researched questions in altmetrics research, however, has been to understand the relationship between altmetrics and citations. There are several studies conducted on data from various social media platforms as well as academic social networks, with the objective of understanding the relationship between altmetrics and citations. Among the initial studies is the work by Eysenbach (2011), in which Twitter data for publications in the Journal of Medical Internet Research were analyzed and it was found that tweets could be an early predictor of citation counts in the domain of medical sciences. A set of 286 research articles constituted the data for analysis and immediate tweets were analyzed. Shema, Bar-IIan, & Thelwall (2014) analyzed two different sized journal samples (12 & 19) for two different time periods (2009 & 2010, respectively) and found that blog citations correlated with citation counts for some journals. Another study in the same year by Haustein et al. (2014) analyzed 1.4 million biomedicine articles indexed in PubMed and Web of Science, published during 2010 and 2012 in more than 5000 bio-medical journals, and found that tweets have weak correlations with citations. They further found that less than 10% of articles from PubMed were found to have a tweet in the collected dataset.

Several other studies over the time also worked on understanding the relationship between altmetrics and citations. Sotudeh, Mazarei, & Mirzabeigi (2015) worked with CiteULike bookmarks for Library & Information Science (LIS) papers and found that they correlate well with citations. Out of 83 source titles examined between 2004 and 2012, ~78% titles had at least one article bookmarked in CitedULike. The citations of LIS papers were found to be weakly but positively correlated with CitedULike bookmarks. Costas, Zahedi, & Wouters (2015a) analyzed a dataset collected for the period of 2005-2012 and confirmed the claims of positive correlations between altmetrics and citations, but found that these correlations are very weak. They also found that only 15-24% of articles get mentioned in social media platforms. Maggio et al.(2018) worked on data of the health professional education (HPE) domain and observed that altmetric indicators are not necessarily found to predict higher citation, but they can help to generate more citations and visibility. They evaluated around 2500 HPE articles published during 2013-2015 in the study. They have also shown that blogging was the most impactful alternative medium of propagating citations. Thelwall (2018) worked on Mendeley reads and found that Mendeley reads have a high correlation with citation counts for ten selected disciplines. In another work, Thelwall & Nevill (2018) examined 27 research areas classified by Scopus and analyzed the data for 10,000 selected articles. They found positive correlations between Mendeley reads and citation counts.

Another interesting direction of research in altmetrics, which perhaps is more closely related to the broader theme of present work, has been to identify various factors or article characteristics that may be positively associated with higher altmetric attention of the research articles. Different factors like journal impact factor, authorship structure, length of article, international collaboration, open access etc. have been analyzed for this purpose. One of the first work (Haustein et al., 2015) on the theme was focused on studying the effect of document properties (mainly discipline, document type, title length, number of pages and references) on social media metrics of scholarly articles. They found that collaboration leads to higher impact. However, they observed that different article characteristics are not similarly associated with bibliometric as well as altmetric impact. For example, they found that the higher impact factor of a journal is positively related to the bibliometric impact of its papers, but it may not be similar for altmetric impact. They have concluded that factors driving social media and citations are different.

Another study (Didegah et al., 2018) observed the differences in the role of selected factors on citations and altmetrics by analyzing a sample of Finnish articles. They found a positive relationship between journal impact factor and bibliometric & altmetric impact, particularly on some platforms. The collaboration was also found to have a positive role in both kinds of impacts. In another work in the same year, Zhang & Wang (2018) observed that in the case of biology, journals with higher impact factor values are not found tweeted often, i.e. correlations are not seen. Regarding studies on the association of open access (OA) with social media attention, some studies (Davis et al., 2008; Gargouri et al., 2010; Norris et al., 2008) studied its association with bibliometric impact, whereas several studies (Poplašen & Grgić, 2017; Wang et al., 2015) tried to look at its association with altmetric impact and found a decisive advantage for OA articles. In general, all the studies concluded that open access journals/ articles get higher bibliometric as well as altmetric impact. One recent study (Holmberg et al., 2020) used data for Finnish universities and showed that there are significant disciplinary and platform-specific differences in the "OA advantage". They found that articles in OA journals in disciplines like veterinary sciences, social & economic geography and psychology receive more citations and attention on social media platforms, while the opposite was found for articles in OA journals within medicine and health sciences.

The previous studies on factors and article characteristics, thus, could help in understanding the impact that these factors or characteristics may have on the altmetric attention of research articles. It was observed that international collaboration and open access are positively associated with higher social media attention of scholarly articles. The journal impact factor, however, was not found to be positively associated with higher social media attention in all cases. The authorship structure's association with social media attention of articles, however, more or less remains unexplored. Most studies on authorship structure essentially targeted international collaboration of authors and the effect of the number of authors on altmetric impact of articles is not very well studied.

Despite existence of several studies on the impact of different factors/ article characteristics on altmetric impact of articles; the association of presence/ absence of social media plugins in journal webpages on the altmetric impact of their articles, has remained totally unexplored. As a result, it is not clear whether the presence of social media plugins in journal webpages could have a positive impact on social media attention of research articles in those journals. To the best of our knowledge, there is no previous study to have studied this question. As a practical consequence, it is also not known whether integration of social media plugin in journal webpage can have a positive impact on altmetric attention or not. This article aims to bridge this gap in knowledge by analyzing a sufficiently large sized data sample.

**Data**

The data for analysis was obtained from two sets of journals, those with social media plugin integrated in their webpage and those that do not have such integration. Two data sources have been used to obtain publication and social media data. The research publication data of two sets of 50 journals each, as described ahead, for the year 2016 was obtained from the Web of Science database. The social media data for each research publication was obtained from Altmetric.com for three social platforms- Twitter, Facebook and Mendeley.

The first set of journals comprised of 50journalsthat have social media plugins integrated in their webpages, for at least two social platforms (out of three platforms- Twitter, Facebook &Mendeley- analyzed by us). The second set comprised of 50journals that do not have plugins integrated for any of the three social platforms. We looked at about 300 journals from different publishers in the process, finally selecting only those that met the criteria of presence or absence of social media plugins in their webpages. In the process of selecting journals, we tried to have a suitable representation of different major publishers as well as disciplines. These two sets are, hereafter, referred to as SMP and NSMP, where SMP refers to set of journals having social media plugin, and NSMP to journals that do not have social media plugins. The SMP set comprised of journals from nine different publishers, namely Elsevier (9 journals), Wiley (9 journals), Oxford Academic (8 journals), Taylor and Francis (7 journals), Nature (5 journals), CellPress (4 journals), PLoS (4 journals), SAGE (2 journals) and ACS publications (2 journals). The NSMP set comprised of journals mainly from four publishers- Springer (16 journals), IEEE (19 journals), and Hindawi (8 journals), and Emerald (7 journals). These journals taken together represent all major disciplines of research. This was an important thing for us to ensure, as it has been seen in earlier studies that articles in some disciplines get slightly higher social media attention than average. Therefore, having representation of all major disciplines will help in neutralizing any extraneous effect of discipline.

After selecting the journals for both SMP and NSMP sets, the publication records for each journal for the year 2016 were downloaded from the Web of Science database. The 50journals in the SMP set, taken together, account for 78,633research publications and the 50journals in the NSMP set, taken together, account for 21,949research publications. All kinds of research publications (different document types in the Web of Science database) were included in the data for analysis. Since all the articles did not have a Digital Object Identifier (DOI), which was the essential linking factor with altmetric data, we had to drop publication records without DOI. Removing publication records without DOI left us with 77,820 articles (with DOI) in the SMP set and 21,929articles (with DOI) in the NSMP set. The analytical study, thus, used a sample of total 99,749 articles drawn from 100 different journals representing 13different publishers. Tables 1 and 2 present the detailed data about the journals used in the analysis.

For each publication record, as obtained above, having a DOI, the social media data corresponding to three platforms- Twitter, Facebook and Mendeley- was obtained from popular altmetric aggregator- Altmetric.com. ADOI lookup was performed in Altmetric.com explorer to get the coverage and mention data for publication records for the three platforms. Out of 77,820 publication records in SMP set, 50,084 publication records had some mentions in Twitter; 13,457 had some mentions in Facebook; and 29,117 had some mentions in Mendeley. Similarly, out of 21,929 publication records in NSMP set, 4,677 publication records had some mentions in Twitter; 1,050 had some mentions in Facebook; and 7,791 had some mentions in Mendeley.

**Methodology**

The social media coverage and mention data for research publications/ articles from the two sets of journals were analyzed computationally by writing Python programs. For each journal, the coverage percentage of its articles in all the three social media platforms is obtained by identifying the proportion of articles that get some mentions in respective platforms. The coverage values for all articles in a journal are averaged. These average values of coverage for all the journals in a set are then aggregated by computing the mean and median of coverage percentage values. This is then done for the NSMP set as well. In addition to coverage percentage, mentions per paper is also computed for each journal for all the three platforms. For example, in order to compute average mentions per paper in the Facebook platform for a journal, the total number of mentions for all the articles in the journal are divided by total number of articles in that journal. This way, the average mentions per paper is computed for all the journals in the two sets- SMP and NSMP. The journal-wise average mention values in each set are then aggregated into a single value summary by computing the mean and median values. This single value summary is taken as a representative value for each of the sets, SMP and NSMP. Thus, coverage and average mentions are the two main parameters computed to distinguish between SMP and NSMP sets and to observe whether SMP set gets higher coverage and average mentions as compared to NSMP set.

In order to understand whether the average mentions per paper are suitable to differentiate between SMP and NSMP journals or not, we employed *area under curve* (AUC) values as a measure. We plotted *receiveroperating characteristic* (ROC) curves to identify the association between individual mentionsof articles for each platform (Twitter, FB, Mendeley) and journal class (SMP or NSMP). Primarily, ROC-AUC procedure is used to test/ compare the performance of one or more classification algorithms, but we have utilized this mechanism to analyze the association strength of features and classes as this was found useful earlier (Hassan et al., 2017). The Logistic Regression model is used to classify the articles published in collected journals. For classification purposes, all publications in the SMP and NSMP sets are tagged with 1 and 0, respectively. The classification task is performed using the built-in Logistic Regression model of sklearn.linear_model module available in python. For each of the three platforms, publication records are classified separately based on their respective mention value to measure the discriminative capability of that platform. Two important parameters involved in the ROC-AUC mechanism are Sensitivity (True Positive Rate or TPR) and, Specificity (1-False Positive Rate or 1-FPR)[2]. These terms are defined as follows

$$Sensitivity = \frac{TP}{TP + FN}$$

$$Specificity = \frac{TN}{TN + FP}$$

where,

TP = True Positive (actual and predicted, both values are positive), FP = False Positive (actual value is negative but prediction value is positive) , TN = True Negative (actual and predicted, both values are negative), and FN = False Negative (actual value is positive but prediction value is negative)

Various parameters of ROC curves are computed using two functions, present in the *sklearn.metrics* module of python. The values of true positive rate, false-positive rate and thresholds are computed using roc_curve function, whereas the AUC measure is found using the roc_auc_score function. We have used *plt* function of a python library *matplotlib.pyplot* to plot the curve with the previously computed parameters of ROC.

---

[2] https://towardsdatascience.com/understanding-auc-roc-curve-68b2303cc9c5. Accessed(10/01/2020)

ROC curves are plotted for each of the three platforms separately, to identify their discrimination strength. Higher the association between mentions and journal class, higher is the distinction capability. An AUC, which is close to 1, shows good separability of the algorithm under consideration as well as the association between the discriminating feature and the target class. In order to find the extent of such relationship, a logistic regression model was deployed, as detailed above. Tweets, Facebook mentions, and Mendeley reads are taken as independent variables while the class label is the dependent variable. The AUC computations confirmed the discriminative nature of mentions for the given task and hence results are found to be useful for the question analyzed.

**Results**

The social media attention levels of all the articles in the two data sets were analyzed to identify if there exist identifiable differences in the coverage and attention levels between articles published in journals having social media plugins (SMP set) and journals that do not have the plugins (NSMP set).

*Differences in overall social media coverage of journals in SMP and NSMP sets*

The first observation is with respect to coverage of articles in journals of the two sets. Tables 1 and 2 show the overall social media coverage levels of articles, as per the data obtained from Altmetric.com, for the SMP and NSMP sets, respectively. The coverage percentage here implies the proportion of articles from a journal that get some social media activity. It can be observed that in the SMP set, 35 out of 50 journals have more than 50% articles covered in the Altmetric.com, whereas in the NSMP set, only 14 out of 50 journals have more than 50% coverage. The mean coverage percentage for the SMP set is 71.556%, whereas the mean coverage percentage for NSMP set is 36.677%. Further, the median of coverage percentage for the SMP set is 85.141whereas, for the NSMP set is 33.486, which is significantly lower. These observations indicate that, in general, those journals that have social media plugins integrated in their webpages get higher proportion of their articles covered in social media platforms. It is interesting to find that several very well-recognized journals like IEEE Transactions on Power Systems, Journal of Mathematical Biology, IEEE Transactions on Industrial Electronics etc. have lower overall social media attention to their articles. The absence of social media plugins in their webpages may be one of the primary reasons for this pattern.

In order to have a more detailed understanding and clear presentation of the results, the coverage levels were clubbed in 5 groups of 20% each. Figure 1 shows the clubbed coverage levels plotted for journals from both the SMP and NSMP sets, figure on the left for the SMP set and figure on the right for the NSMP set. The x-axis represents the coverage levels and the y-axis represents the number of journals (frequency in that coverage level). Five different bars represent the frequency of each coverage level, in each figure. It is observed that, in the case of SMP set, there are more than 25 journals (which is more than 50% of SMP set) grouped in the range of 80%-100% coverage levels in different social media platforms. Only three journals are having coverage levels less than equal to 20% in the SMP set. In contrast, most of the journals in NSMP set (precisely 27 out of 50 journals) are having coverage levels less than 40%. Only four journals are found to be in a higher coverage level group (80%-100%), which is merely 8% of the whole NSMP set. These observations provide a clear indication that the journals with social media plugin integrated tend to have more social media attention to their articles. The integration of social media plugins in journal webpages is, thus, observed to have a clearly identifiable advantage in terms of social media coverage.

It would also be important here to look at any possible interaction of the above noted patterns with disciplinary variations in social media coverage of articles. Several previous studies (such as Banshal et al., 2019a; 2019b) have shown that articles in some disciplines tend to get higher social media coverage as compared to other disciplines. Articles in disciplines like medical and biological sciences have been found to get higher social media coverage and attention as compared to several other disciplines like Information Science, Mathematics, Engineering etc. We, therefore, look at the disciplinary representation of journals in the two sets and their social media coverage levels. The SMP set comprises of journals from all major disciplines like, Medical Sciences, Biological Sciences, Chemistry, Material Sciences, Mathematics, Earth Sciences, Agricultural Sciences, Environmental Sciences, Engineering, and Multidisciplinary. The NSMP set also has journals representing all major disciplines.

If we look at some specific examples, we could understand it more clearly that social media plugin integration is the main differentiating factor across the journals in the same discipline. For example, in SMP set, "Journal of Infectious Diseases", a journal from Medical Sciences, has coverage percentage of 88.42%, whereas another journal, "Biomed Research International", from the same discipline, in NSMP set has coverage percentage of 41.31%. Another example could be from the Material Science area, where it is observed that journals, "ACS Nano", "Materials Today", "Small", all in SMP set, get coverage percentages of 75%, 69% and 46%, respectively. On the other hand, a journal "Rapid Prototyping Journal", again in the Material Science area, in NSMP set has coverage percentage of 4.5%. There are several other examples of similar kind that substantiate the fact that social media plugin presence is a major differentiator in coverage across the journals in the same discipline.

### *Journals with and without Twitter plugin*

After analyzing the overall coverage levels, we analyzed the platform-wise coverage and average mention statistics. The first analysis was for the Twitter data of journals in SMP and NSMP sets. Here, SMP represents the journals that have Twitter plugin integrated and NSMP represents journals that do not have Twitter plugin integrated in their webpages. Tables 3 and 4 present the coverage percentage and tweets per paper for journals in SMP and NSMP sets, respectively. It can be observed here that the mean coverage percentage of articles in Twitter platform for SMP set is 62.3%, which is more than three times higher than the mean coverage percentage of NSMP set, which is 21.71%. Further, the average median values of coverage percentages for the SMP and NSMP sets, for Twitter platform data, are 73.25% and 14.02%, respectively.  It can be said that average coverage percentage, for Twitter platform data, for SMP and NSMP sets are significantly different.  If we look at individual journals, we observe that in the SMP set, there are 18 journals (out of 50) that have coverage percentage of more than 90%. About 60% journals in the SMP set have coverage percentage higher than 60%. On the other hand, in the NSMP set, only 12 journals (out of 50) are found to have a good coverage percentage. In addition to the coverage percentage, the average tweets per paper was also computed for each journal in the two sets. Average tweets per paper value for a journal is computed by dividing total tweets for articles in that journal by the number of articles. These journal-wise average tweet values are then aggregated for the SMP and NSMP sets. It is observed that the SMP set has mean tweets per paper value as 14.135, whereas the NSMP set has a mean of tweets per paper value as 2.925. The median values of tweets per paper for the SMP and NSMP sets are 6.515 and 2.355, respectively. Some specific examples of journals to look may be Proceedings of the IEEE, IEEE Transactions on Pattern Analysis and Machine

Intelligence, IEEE Transactions on Industrial Electronics, which all, despite being very well-recognized journals, have quite low Twitter coverage and average tweet per paper values being 10.795, 2.504 and 1.267, respectively. Integration of Twitter plugin in journal webpages thus seems to have a clear advantage here.

To further understand the differentiation between the whole data of SMP and NSMP sets, a comparative visualization is created with the average tweets per paper values in the two sets. Figure 2 presents the plot of average tweets per paper value for all the journals in the two sets. Here, the x-axis represents a journal and the y-axis represents tweet per paper value for that journal. For observing a clear differentiation, journals on the x-axis are ordered (from left to right) in descending order of their tweet per paper value. The two curves plot average tweets per paper value for different journals in the two sets. It is clearly observed that journals in the SMP set get higher values for tweets per paper as compared to journals in the NSMP set. The curves indicate that a larger number of journals in the SMP set has significantly higher average tweet per paper value as compared to NSMP set.

Since we used average tweet per paper value as a differentiating feature across the two sets (classes) of journals, we analyzed the data at a finer granularity of tweets for individual articles. To measure the degree of association between tweets and journal class, a ROC curve is plotted by taking tweets of publications in both sets as the independent variable. Figure 3 shows the ROC curve for the tweet data for articles in journals in the two sets. It is observed that AUC value is 0.730, which shows that a large area is covered under the curve. The large AUC shows a strong association between the tweets and journal classes. A value of AUC = 0.725 in an earlier study by Hassan et al., (2017) used a similar argument. This AUC value thus confirms the association between tweets per paper and journals classes and serves as evidence that average tweets per paper would also have a strong association, and hence can be used to differentiate articles in journals in SMP and NSMP sets.

The discipline connection in the case of Twitter platform is also analyzed. We compared the coverage percentage and average tweets per paper for different journals in the two sets. Some interesting example to list would be, journal "Nature Communications" in SMP set, with average tweet per paper value of 24.459, as compared to journal "Complexity" in NSMP set, having average tweet per paper value as 9.444. Both of these journals represent the same disciplinary area. Another example could be journal "Plant, Cell and Environment" in SMP set, with Twitter coverage of 35.455% and average tweet per paper as 4.167, as against journal "British Food Journal" in NSMP set, having Twitter coverage of 12.766% and average tweet per paper as 3.375. Both these journals are from the Agricultural Sciences area. Thus, in the case of Twitter platform data as well, integration of plugin shows a clear differentiation in coverage and mentions per paper, for journals from the same discipline.

*Journals with and without Facebook plugin*

The Facebook platform data was the next to be analyzed. Tables 5 and 6 present the results for mean coverage percentage and mentions per paper value for the Facebook platform for SMP and NSMP sets, respectively. It is observed that the mean coverage percentage for SMP set is 17.942%, whereas for the NSMP set it is 7.296%. The median of coverage percentage for SMP and NSMP sets are 11.498% and 0.973%, respectively. In terms of Facebook mentions per paper, we have observed that the mean mentions per paper value for SMP set is 1.948and for NSMP set it is 1.001. The median values of mentions per paper for SMP and NSMP sets are 1.462 and 1, respectively. We observe that, in the SMP set, there are 22 journals having coverage percentage of more than 10%, as against 5 journals in NSMP set. There are 15 journals in NSMP set that have coverage percentage in the Facebook platform as 0.

In order to clearly visualize the overall pattern of average Facebook mentions per paper, we create plot with journals on the x-axis and average Facebook mention per paper for that journal on the y-axis. Figure 4 shows the two curves for SMP and NSMP set of journals, with journals on the x-axis arranged in descending order of their average Facebook mention per paper for each journal. A reasonable difference between the average mentions in the two sets can be observed in the plot, with SMP set getting higher values as compared to NSMP set. The peak value for the SMP set is higher than 10 as compared to peak value for the NSMP set around 4. Thus, it is observed that the average mentions per paper value for the journals in the two sets vary, with SMP having overall higher values as compared to the NSMP set.

In order to test for the discriminative capacity of average mentions per paper across the two sets of journals, a ROC curve for Facebook platform data is plotted, in figure 5. It is observed that AUC value is 0.555, which is not significantly high. Thus, Facebook mentions per paper has a weak association with the classification of journals in the two sets. The AUC value suggests that average FB mentions are not as encouraging as that of tweets per paper. Relatively low use of Facebook by academicians from several regions of the World could be a reason for this (Hank et al., 2014). The lesser magnitude of differences in Facebook mentions per paper in the two sets shows a similar pattern.

The discipline connection has also been analyzed for Facebook data. The journals in SMP and NSMP sets are compared in terms of coverage levels and average mentions for several disciplines. Though the magnitude of differences is not high in case of the Facebook platform, still some positive advantage is seen of having Facebook plugin integrated in journals. For example, the journal "Cell", in SMP set, has a coverage percentage of 60.98% and average mentions per paper as 4.345, as against the journal "Parasitology Research", in NSMP set, with coverage percentage as 6.767% and average mentions per paper as 1.556. There are several other examples of a slight advantage of integrating Facebook plugin in journals, of the same discipline, in terms of coverage percentage and mentions per paper.

Thus, perusal of all the results for Facebook platform show that in the case of this platform too, there is an observed advantage of integrating plugin in journal webpages on their coverage and mentions per paper value. However, here the magnitude of differences is not very high between the journals in the two sets.

### *Journals with and without Mendeley plugin*

The third platform for which data was analyzed was Mendeley. In this case, unlike the other two platforms, we found only 22 journals having Mendeley plugin integrated. Thus, in this case, the NSMP set comprises of 78 journals. Tables 7 and 8 show the results for mean coverage percentage and mentions per paper for SMP and NSMP sets, respectively. It is observed that the mean coverage percentage for SMP and NSMP sets are 84.927% and 36.256%. Similarly, the median values for coverage percentage for SMP and NSMP sets are 94.252% and 32.36%, respectively. Thus, a significant value of difference in coverage percentage is seen in journals and SMP and NSMP sets.

The mean and median values for mentions per paper was also computed for both sets. It is observed that for SMP set the mean of mentions per paper is 49.656 and the median is 50.971, whereas for NSMP set mean and median of mentions per paper are37.002 and24.899. Though, the mean value of mentions per paper of the NSMP set is close to the SMP set, the median value of the SMP set is higher than two times of the NSMP set.

The spread of average reads per paper for the journals across the two sets (SMP & NSMP) has been plotted in figure 6, for a better visualization. Here too, the x-axis has journals arranged in descending order of Mendeley reads per paper and y-axis value represents the Mendeley reads per paper for that journal. The two curves show the patterns for SMP and NSMP sets. The difference between the average reads in the SMP and NSMP sets is quite high as compared to the variation in Twitter and Facebook platforms. Though the highest value from both the sets lies around 200 but the overall difference between the averages is evidently noticeable in trend lines from the figure. The tail trend of the NSMP set is quite lower than the SMP set. The larger size of the NSMP set could be a reason for a relatively much higher difference in this case. Nevertheless, it is observed that journals with Mendeley plugin get much higher reads per paper as compared to journals that do not have such plugin.

The discriminative capacity of average Mendeley reads is also analyzed through the ROC plot by looking at article level read statistics. Figure 7 shows the ROC curve, representing the relationship between journal class and Mendeley reads. The AUC value in this case is 0.673, which could be taken as an acceptable value for the discriminative capacity of Mendeley reads between the two classes. This value thus indicates the validity of representing the average reads as the differentiating indicator between SMP and NSMP journals and confirms its use in the analysis.

The coverage and attention levels of the Mendeley platform have also been analyzed with respect to discipline connection. In the case of Mendeley platform data too, there are several examples of journals in the same discipline having differential coverage and average reads per paper value owing to Mendeley plugin integration. One example could be a journal, "Journal of The American Statistical Association", in the SMP set, that has coverage of 95.031% and average reads per paper as 24.353. Another journal "Acta Mathematica Sinica", in NSMP set from the similar area, has 5.172 % coverage and 6.5 average reads per paper. The Medical Sciences journal "PLoS Neglected Tropical Diseases", in the SMP set has coverage 97.603% and average reads per paper 55.24 as against to the journal "Disease Markers", in the NSMP set with coverage 41.615% and the average reads per paper as 22.687. Thus, the plugin advantage of social media attention is observed in Mendeley platform too, as observed in journals from the same discipline in the two sets. The difference in magnitudes here, however, is not as significant as Twitter.

**Conclusion**

The paper tried to answer the question, whether the presence of social media plugin in journal webpages result in higher social media visibility of its articles. Two data samples of 50 journals each, with and without social media plugin, are studied and the social media activity around their publications is captured. Social media data of a total of 99,749 articles, from these 100 journals, were obtained and analyzed for the purpose. The journals in the data sample represent different disciplines as well as different publishers. The coverage and mentions per paper values in three platforms- Twitter, Facebook and Mendeley- are analyzed. The ROC curves are also plotted and AUC computed for mentions per paper data for various platforms to understand the discriminative power of mentions per paper values for the two journal sets.

Results show a clearly identifiable differentiation between social media coverage and mention per paper for articles published in a journal with and without social media plugins. A journal that has a plugin for a social platform integrated in its website is found to attract higher coverage and attention to its articles in that social media platform. This is found to be valid for all the three social media platforms analyzed, though the degree of differentiation varies across the platforms. The impact of the presence of plugin is seen most clearly and notably for the Twitter platform. The journals that have Twitter plugin integrated in their websites get much higher coverage and attention of their articles in Twitter. Mendeley platform also shows differentiation in coverage and mentions per paper between the journals, with and without Mendeley plugins. The Facebook plugin integration in journals also shows a positive impact, though the magnitude of difference in impact on social media coverage and attention is lesser as compared to other platforms.

The ROC plots made and the AUC values computed for article-level mentions per paper data, for the three platforms, show that average mentions per paper is a suitable discriminator of the two sets of journals, those with the plugin and those without them. Twitter mention data has the highest value of AUC followed by Mendeley. Facebook platform mention data shows relatively lower AUC value. The values indicate higher differentiation in mentions per paper values in the Twitter platform as compared to Facebook.

The set of journals analyzed not only represented different publishers but also different disciplines. The social media data of articles is, thus, a suitable representation of various disciplines. Given the fact that some disciplines are known to attract higher social media attention, it was important to see that the positive differentiation of the presence of social media plugin on social media coverage and attention holds for journals in different disciplines. The results show several examples of journals from the same disciplinary area, in SMP and NSMP sets, where journals in the SMP set get much higher social media coverage and attention of their articles as compared to journals in NSMP set. This confirms that the impact of the presence of social media plugin in journal webpages is independent of discipline.

The different results for overall coverage, mean and median values of coverage percentages, mentions per paper values and the AUC values computed for journals in two sets for different platforms confirm that presence of social media plugin in journal webpages give higher social media coverage and attention to articles published in that journal. The results, therefore, provide a positive answer to the research question proposed through conclusive evidence from three social platforms.

It would be important to try to understand possible reasons for these patterns observed and to look at what practical implications it may have for scholarly publishing. It appears that the presence of social media plugin on journal webpages acts as a kind of facilitator for social media sharing and mentioning. Perhaps, authors and readers prefer to click the readily available plugin links on the webpages to either share the article in their profiles or to write their comments/ opinions about the article. These clicks and shares may be originating from different kinds of browsing sessions. First, authors of articles in the journal may find it easier and more convenient to use the plugins available at the journal webpage to disseminate and share their articles. Secondly, the readers of a journal may also find it useful to use the plugin links to share new findings reported in the articles with their peers and students. Both these activities result in higher social shares & mentions of articles published in journals that provide readily available sharing/ commenting mechanism in the form of plugins. However, journals that do not provide the plugins integrated in their webpages are at a disadvantage in terms of ease and convenience of sharing in social media platforms.

These results have very important practical implications for journal publishers. Results indicate that publishing houses should actively provide social media plugin integration in their journal webpages to increase the social media visibility (altmetric impact) of their articles. Given the fact that altmetric values correlate well with citation counts of articles, this indirectly implies that integration of social media plugins in journal webpage may also indirectly facilitate in higher citation impact of articles in the journal. Journals that do not integrate the social media plugin may continue to be at a disadvantage in respect to social media sharing/ mentions of their articles.

**Acknowledgements**

The authors would like to acknowledge the support of Stacy Konkiel, Director of Research Relations at Digital Science, for providing access to Altmetric.com data.

# FIGURES

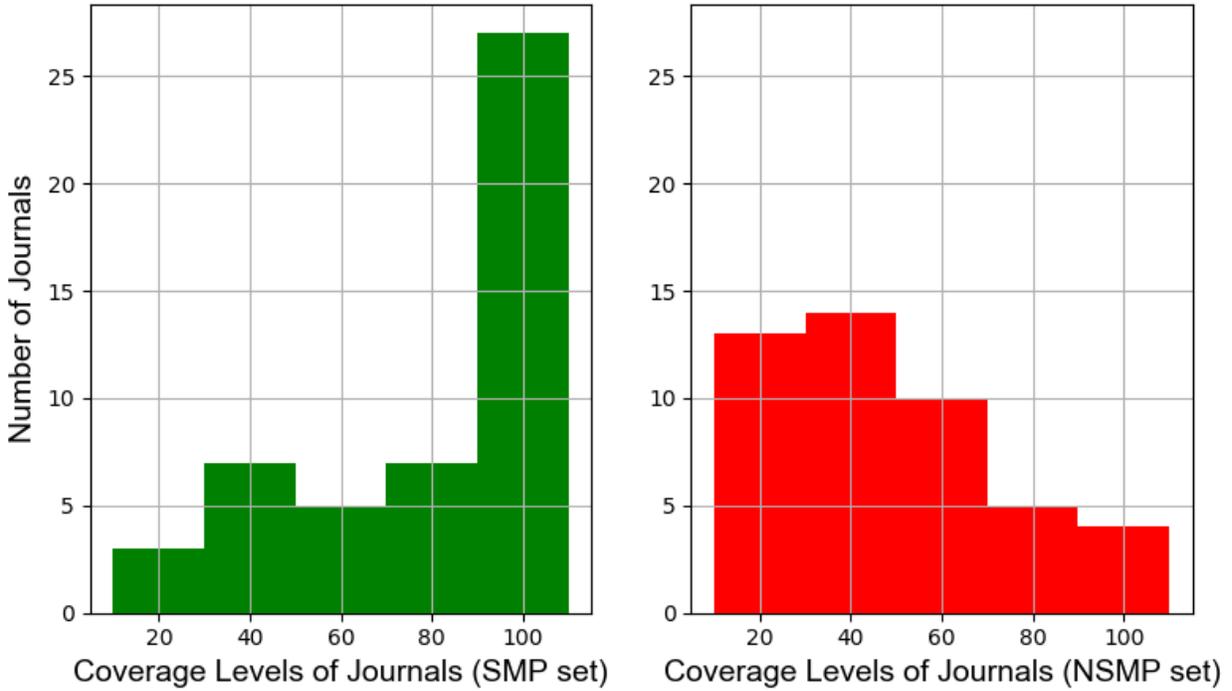

Figure 1: Coverage Levels of SMP & NSMP sets of journals

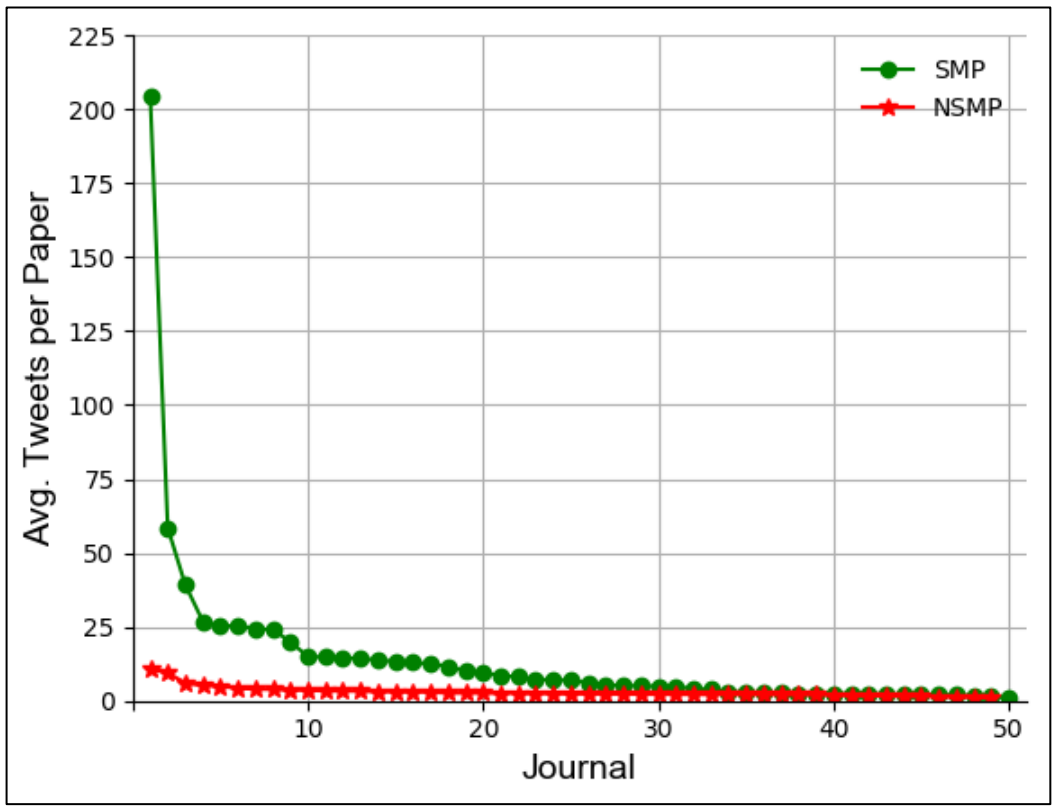

Figure 2: Comparison of average Tweets per paper for journals in SMP & NSMP sets

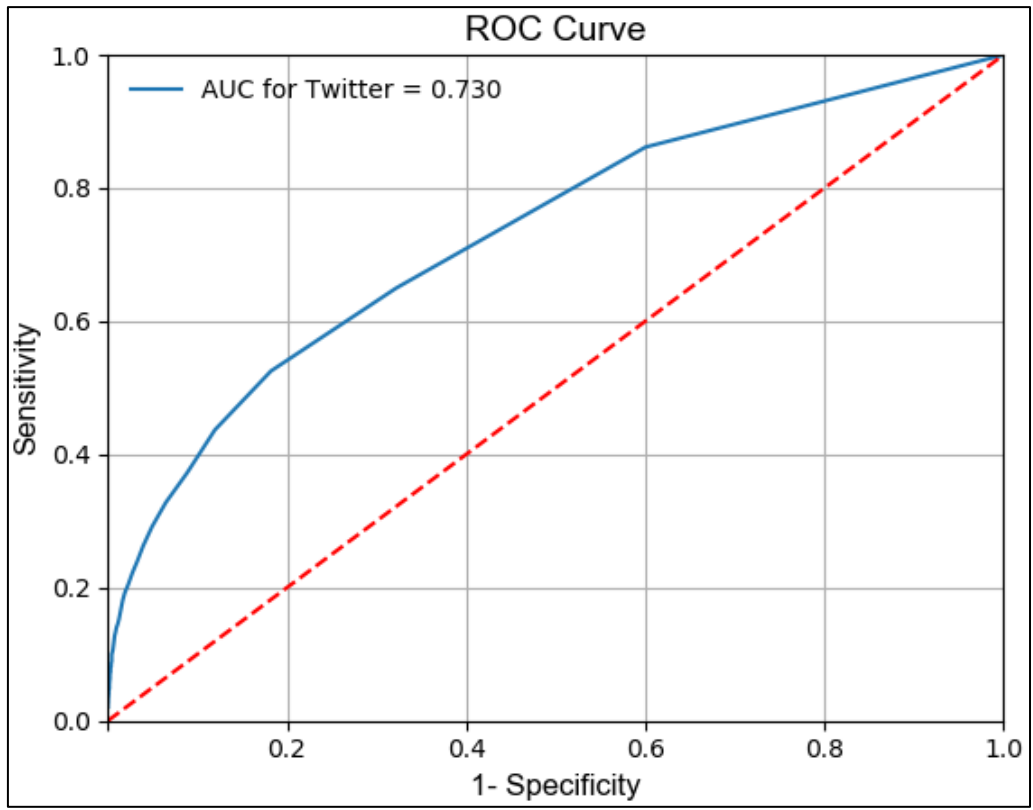

Figure 3: ROC curve for average Tweets to classify papers in SMP & NSMP sets

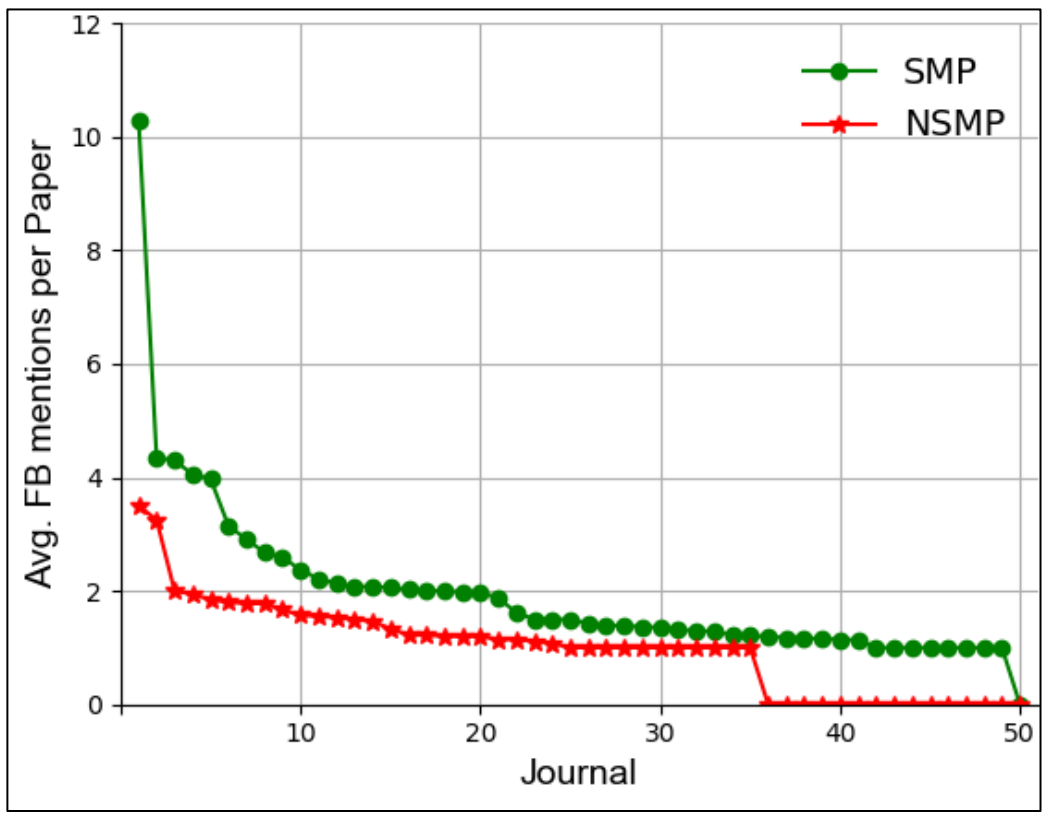

Figure 4: Comparison of average FB mentions per paper for journals in SMP & NSMP sets

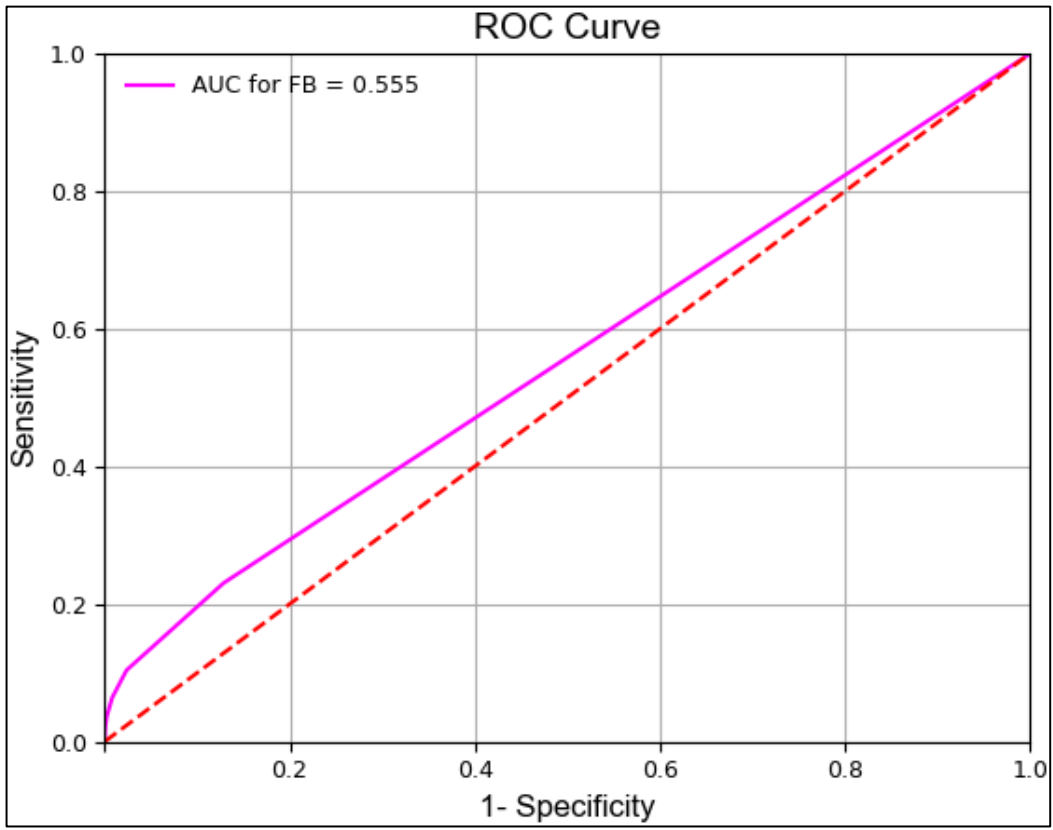

Figure 5: ROC Curve for average FB mentions to classify papers in SMP & NSMP sets

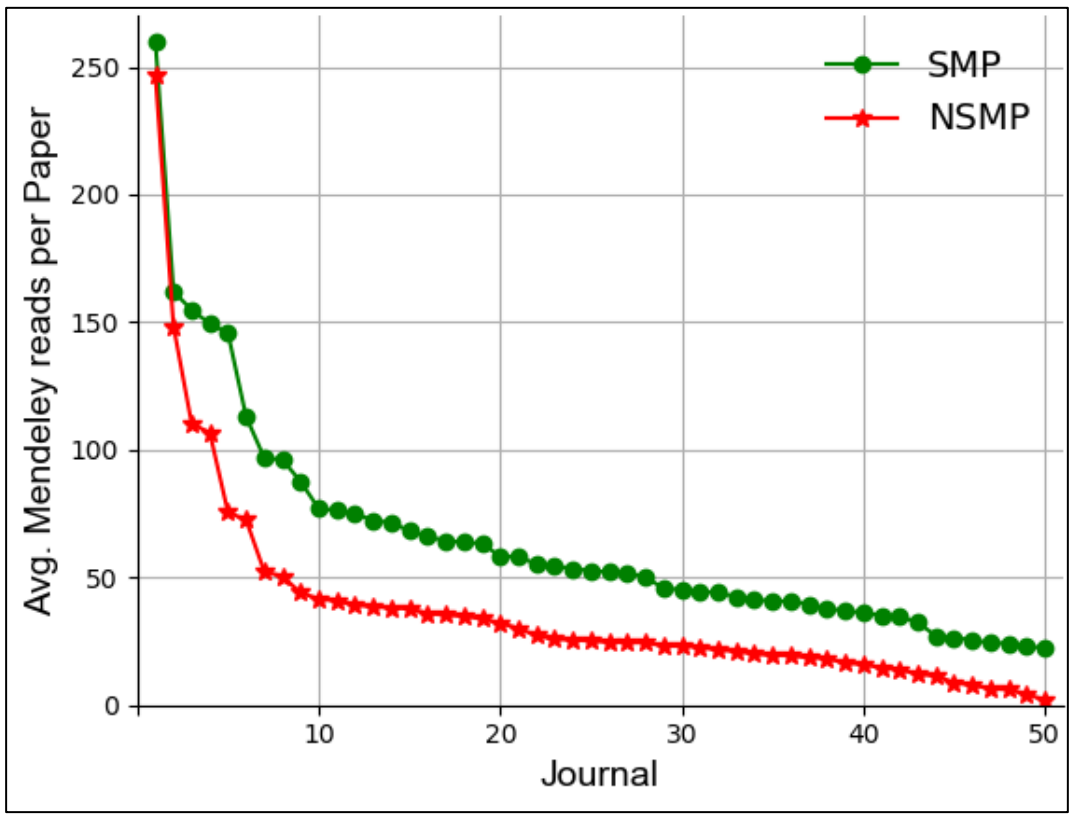

Figure 6: Comparison of average Mendeley reads per paper for journals in SMP & NSMP sets

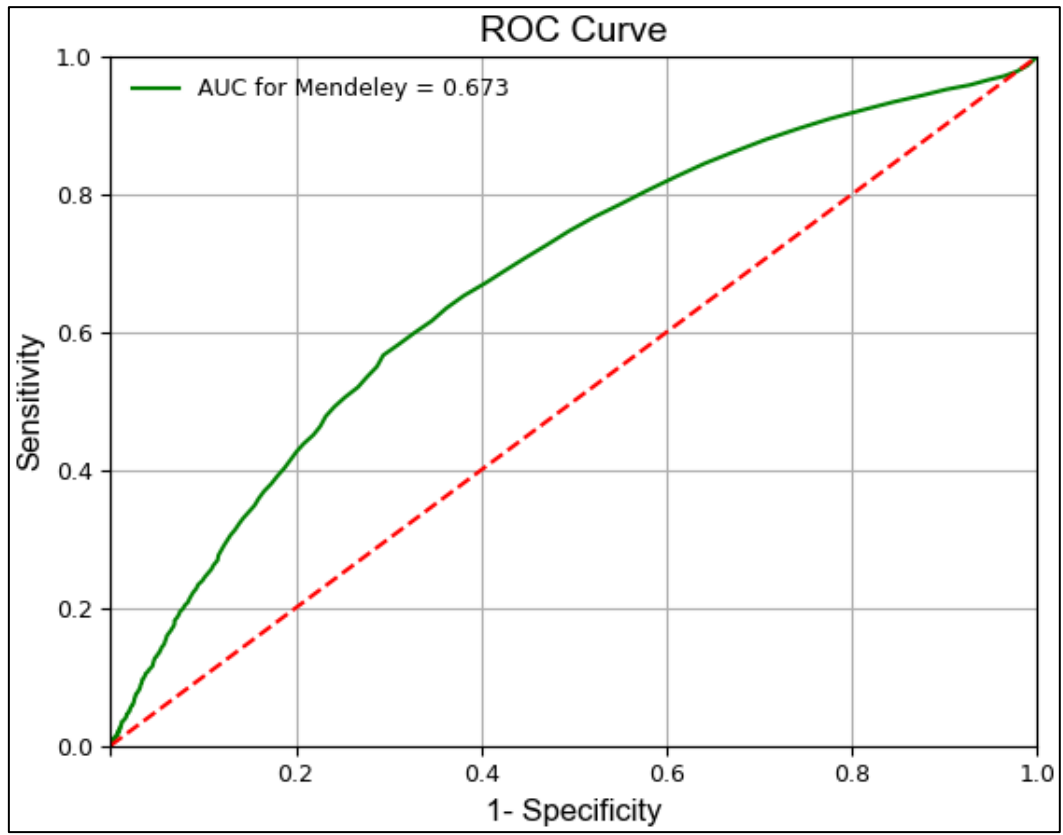

Figure 7: ROC Curve for average Mendeley reads to classify papers in SMP & NSMP sets

# TABLES

**Table1: Publisher and Coverage Summary for journals with social media plugin**

| S. No. | Journal Name | Publisher | WoS Records | Records with DOI | Records found in altmetric.com | Coverage Level % | Twitter | FB | Mendeley |
|---|---|---|---|---|---|---|---|---|---|
| 1 | Acs Nano | Acs Publications | 1,285 | 1,285 | 975 | 75.875 | YES | YES | YES |
| 2 | Advanced Energy Materials | Wiley | 420 | 419 | 94 | 22.434 | YES | YES | NO |
| 3 | Advanced Functional Materials | Wiley | 881 | 881 | 322 | 36.549 | YES | YES | NO |
| 4 | Advanced Materials | Wiley | 1,172 | 1,169 | 70 | 5.988 | YES | YES | NO |
| 5 | American Journal of Sports Medicine | Sage | 438 | 438 | 388 | 88.584 | YES | YES | YES |
| 6 | Annals of Botany | Oxford University Press | 217 | 209 | 208 | 99.522 | YES | YES | YES |
| 7 | Applied Catalysis B Environmental | Elsevier | 789 | 788 | 348 | 44.162 | YES | YES | NO |
| 8 | Bioinformatics | Oxford University Press | 812 | 811 | 809 | 99.753 | YES | YES | YES |
| 9 | Biomaterials | Elsevier | 651 | 651 | 451 | 69.278 | YES | YES | NO |
| 10 | Biophysical Journal | Cell Press | 3,814 | 3,644 | 874 | 23.985 | YES | YES | NO |
| 11 | Biosensors & Bioelectronics | Elsevier | 1,015 | 1,014 | 699 | 68.935 | YES | YES | NO |
| 12 | Briefings in Bioinformatics | Oxford University Press | 93 | 93 | 93 | 100 | YES | YES | YES |
| 13 | Cell Reports | Cell Press | 1,101 | 1,100 | 1,081 | 98.273 | YES | YES | NO |
| 14 | Cell | Cell Press | 690 | 661 | 648 | 98.033 | YES | YES | NO |
| 15 | Chinese Journal of Catalysis | Elsevier | 239 | 239 | 9 | 3.766 | YES | YES | NO |
| 16 | Clinical Infectious Diseases | Oxford University Press | 839 | 777 | 628 | 80.824 | YES | YES | YES |
| 17 | Current Biology | Cell Press | 847 | 846 | 825 | 97.518 | YES | YES | NO |

| # | Journal | Publisher | | | | | | | |
|---|---------|-----------|------|------|------|--------|-----|-----|-----|
| 18 | Database The Journal of Biological Databases And Curation | Oxford University Press | 166 | 164 | 139 | 84.756 | YES | YES | YES |
| 19 | European Journal of Sport Science | Taylor & Francis | 155 | 155 | 153 | 98.71 | YES | YES | YES |
| 20 | Expert Review of Medical Devices | Taylor & Francis | 116 | 116 | 116 | 100 | YES | YES | YES |
| 21 | International Journal of Production Research | Taylor & Francis | 460 | 460 | 118 | 25.652 | YES | YES | YES |
| 22 | International Journal of Remote Sensing | Taylor & Francis | 334 | 334 | 86 | 25.749 | YES | YES | YES |
| 23 | Journal of Catalysis | Elsevier | 358 | 358 | 87 | 24.302 | YES | YES | NO |
| 24 | Journal of Experimental Botany | Oxford University Press | 514 | 514 | 487 | 94.747 | YES | YES | YES |
| 25 | Journal of Infectious Diseases | Oxford University Press | 691 | 691 | 611 | 88.423 | YES | YES | YES |
| 26 | Journal of Power Sources | Elsevier | 1,651 | 1,651 | 320 | 19.382 | YES | YES | NO |
| 27 | Journal of Sports Sciences | Taylor & Francis | 298 | 297 | 292 | 98.316 | YES | YES | YES |
| 28 | Journal of The American Statistical Association | Taylor & Francis | 161 | 161 | 159 | 98.758 | YES | YES | YES |
| 29 | Materials Today | Elsevier | 183 | 96 | 66 | 68.75 | YES | YES | NO |
| 30 | Molecular Ecology | Wiley | 433 | 433 | 424 | 97.921 | YES | YES | NO |
| 31 | Nano Letters | Acs Publications | 1,195 | 1,194 | 993 | 83.166 | YES | YES | YES |
| 32 | Nanotoxicology | Taylor & Francis | 150 | 150 | 93 | 62 | YES | YES | YES |
| 33 | Nature Communications | Nature | 3,687 | 3,685 | 3,452 | 93.677 | YES | YES | NO |
| 34 | Nature Materials | Nature | 291 | 277 | 242 | 87.365 | YES | YES | NO |
| 35 | Nature | Nature | 2,807 | 2,397 | 2,390 | 99.708 | YES | YES | NO |
| 36 | New Phytologist | Wiley | 538 | 530 | 524 | 98.868 | YES | YES | NO |
| 37 | NPG Asia Materials | Nature | 104 | 104 | 98 | 94.231 | YES | YES | NO |
| 38 | Nucleic Acids | Oxford | 1,298 | 1,298 | 1,226 | 94.453 | YES | YES | YES |

| | Research | University Press | | | | | | | |
|---|---|---|---|---|---|---|---|---|---|
| 39 | Plant Biotechnology Journal | Wiley | 203 | 203 | 174 | 85.714 | YES | YES | NO |
| 40 | Plant, Cell And Environment | Wiley | 221 | 220 | 97 | 44.091 | YES | YES | NO |
| 41 | PLoS Computational Biology | PLoS | 582 | 582 | 580 | 99.656 | YES | YES | YES |
| 42 | PLoS Neglected Tropical Diseases | PLoS | 876 | 876 | 856 | 97.717 | YES | YES | YES |
| 43 | PLoS One | PLoS | 23,020 | 23,019 | 19,687 | 85.525 | YES | YES | YES |
| 44 | PLoS Pathogens | PLoS | 685 | 684 | 683 | 99.854 | YES | YES | YES |
| 45 | Progress in Materials Science | Elsevier | 49 | 49 | 24 | 48.98 | YES | YES | NO |
| 46 | Progress in Photovoltaics | Wiley | 143 | 143 | 48 | 33.566 | YES | YES | NO |
| 47 | Progress in Polymer Science | Elsevier | 49 | 49 | 22 | 44.898 | YES | YES | NO |
| 48 | Scientific Reports | Nature | 21,063 | 21,057 | 14,968 | 71.083 | YES | YES | NO |
| 49 | Small | Wiley | 660 | 659 | 300 | 45.524 | YES | YES | NO |
| 50 | Statistical Methods in Medical Research | Sage | 189 | 189 | 130 | 68.783 | YES | YES | NO |
| | | | | | **Mean** | **71.556** | | | |
| | | | | | **Median** | **85.141** | | | |

**Table 2: Publisher and Coverage Summary for journals without social media plugin**

| S. No. | Journal name | Publisher | WoS Records | Records with DOI | Records found in altmetric.com | Coverage Level % | Twitter | FB | Mendeley |
|---|---|---|---|---|---|---|---|---|---|
| 1 | Acta Mathematica Sinica | Springer | 116 | 116 | 8 | 6.897 | NO | NO | NO |
| 2 | Advances in Mathematical Physics | Hindawi | 122 | 122 | 17 | 13.934 | NO | NO | NO |
| 3 | Aircraft Engineering and Aerospace Technology | Emerald | 90 | 90 | 5 | 5.556 | NO | NO | NO |
| 4 | Analytical & Bioanalytical Chemistry | Springer | 837 | 837 | 539 | 64.397 | NO | NO | NO |
| 5 | Annals of Biomedical Engineering | Springer | 306 | 305 | 200 | 65.574 | NO | NO | NO |
| 6 | Annals of Operations Research | Springer | 312 | 312 | 72 | 23.077 | NO | NO | NO |
| 7 | Applied Microbiology & Biotechnology | Springer | 887 | 887 | 523 | 58.963 | NO | NO | NO |
| 8 | Autonomous Robots | Springer | 78 | 78 | 78 | 100 | NO | NO | NO |
| 9 | Biochemistry-Moscow | Springer | 171 | 171 | 46 | 26.901 | NO | NO | NO |
| 10 | Biomed Research International | Hindawi | 1853 | 1852 | 766 | 41.361 | NO | NO | NO |
| 11 | British Food Journal | Emerald | 188 | 188 | 66 | 35.106 | NO | NO | NO |
| 12 | Bulletin of Mathematical Biology | Springer | 95 | 95 | 61 | 64.211 | NO | NO | NO |
| 13 | Canadian Respiratory Journal | Hindawi | 95 | 95 | 50 | 52.632 | NO | NO | NO |
| 14 | Chinese Annals of Mathematics | Springer | 73 | 73 | 1 | 1.37 | NO | NO | NO |
| 15 | Complexity | Hindawi | 194 | 194 | 56 | 28.866 | NO | NO | NO |
| 16 | Disease Markers | Hindawi | 161 | 161 | 67 | 41.615 | NO | NO | NO |
| 17 | Empirical Software Engineering | Springer | 73 | 73 | 68 | 93.151 | NO | NO | NO |
| 18 | IEEE Access | IEEE | 820 | 815 | 23 | 2.822 | NO | NO | NO |
| 19 | IEEE Communications Letters | IEEE | 635 | 635 | 127 | 20 | NO | NO | NO |
| 20 | IEEE Journal on Selected Areas in Communications | IEEE | 301 | 301 | 109 | 36.213 | NO | NO | NO |
| 21 | IEEE Transactions on Antennas and Propagation | IEEE | 677 | 677 | 150 | 22.157 | NO | NO | NO |
| 22 | IEEE Transactions on Biomedical Engineering | IEEE | 269 | 269 | 156 | 57.993 | NO | NO | NO |
| 23 | IEEE Transactions on Communications | IEEE | 405 | 405 | 143 | 35.309 | NO | NO | NO |
| 24 | IEEE Transactions on Geoscience and Remote Sensing | IEEE | 573 | 573 | 147 | 25.654 | NO | NO | NO |
| 25 | IEEE Transactions on Image Processing | IEEE | 446 | 446 | 183 | 41.031 | NO | NO | NO |
| 26 | IEEE Transactions on | IEEE | 766 | 765 | 157 | 20.523 | NO | NO | NO |
| | Industrial Electronics | | | | | | | | |
| 27 | IEEE Transactions on Mobile Computing | IEEE | 225 | 224 | 94 | 41.964 | NO | NO | NO |
| 28 | IEEE Transactions on Neural Networks and Learning Systems | IEEE | 228 | 228 | 92 | 40.351 | NO | NO | NO |
| 29 | IEEE Transactions on Pattern Analysis and Machine Intelligence | IEEE | 193 | 193 | 142 | 73.575 | NO | NO | NO |
| 30 | IEEE Transactions on Power Electronics | IEEE | 736 | 736 | 146 | 19.837 | NO | NO | NO |
| 31 | IEEE Transactions on Power Systems | IEEE | 529 | 528 | 150 | 28.409 | NO | NO | NO |
| 32 | IEEE Transactions on Signal Processing | IEEE | 486 | 486 | 249 | 51.235 | NO | NO | NO |
| 33 | IEEE Transactions on Visualization and Computer Graphics | IEEE | 245 | 240 | 82 | 34.167 | NO | NO | NO |
| 34 | IEEE Transactions on Wireless Communications | IEEE | 631 | 631 | 207 | 32.805 | NO | NO | NO |
| 35 | Industrial Management & Data Systems | Emerald | 96 | 96 | 30 | 31.25 | NO | NO | NO |
| 36 | International Journal of Advanced Manufacturing Technology | Springer | 1420 | 1420 | 317 | 22.324 | NO | NO | NO |
| 37 | International Journal of Numerical Methods for Heat & Fluid Flow | Emerald | 132 | 132 | 11 | 8.333 | NO | NO | NO |
| 38 | Journal of Lightwave Technology | IEEE | 691 | 691 | 158 | 22.865 | NO | NO | NO |
| 39 | Journal of Mathematical Biology | Springer | 133 | 133 | 72 | 54.135 | NO | NO | NO |
| 40 | Kybernetes | Emerald | 101 | 101 | 7 | 6.931 | NO | NO | NO |
| 41 | Multimedia Tools & Applications | Springer | 876 | 876 | 105 | 11.986 | NO | NO | NO |
| 42 | Online Information Review | Emerald | 68 | 68 | 28 | 41.176 | NO | NO | NO |
| 43 | Parasitology Research | Springer | 532 | 532 | 439 | 82.519 | NO | NO | NO |

| S. No. | Journal Name | Publisher | Records with DOI | | Coverage % | | | |
|---|---|---|---|---|---|---|---|---|
| 44 | Proceedings of The IEEE | IEEE | 159 | 155 | 95 | 61.29 | NO | NO | NO |
| 45 | Rapid Prototyping Journal | Emerald | 90 | 90 | 9 | 10 | NO | NO | NO |
| 46 | Scanning | Hindawi | 102 | 102 | 21 | 20.588 | NO | NO | NO |
| 47 | Security and Communication Networks | Hindawi | 475 | 475 | 69 | 14.526 | NO | NO | NO |
| 48 | Shock and Vibration | Hindawi | 390 | 390 | 46 | 11.795 | NO | NO | NO |
| 49 | Springerplus | Springer | 2115 | 2115 | 850 | 40.189 | NO | NO | NO |
| 50 | Surgical Endoscopy & Other Interventional Techniques | Springer | 753 | 752 | 649 | 86.303 | NO | NO | NO |
| | | | | | Mean | 36.677 | | | |
| | | | | | Median | 33.486 | | | |

**Table 3: Mention statistics for journals with Twitter plugin**

| S. No. | Journal Name | Records with DOI | Tweeted | Coverage % | Twitter Mentions | Avg. Twitter Mentions |
|---|---|---|---|---|---|---|
| 1 | Acs Nano | 1,285 | 793 | 61.712 | 3,941 | 4.97 |
| 2 | Advanced Energy Materials | 419 | 48 | 11.456 | 68 | 1.417 |
| 3 | Advanced Functional Materials | 881 | 123 | 13.961 | 327 | 2.659 |
| 4 | Advanced Materials | 1,169 | 41 | 3.507 | 123 | 3 |
| 5 | American Journal of Sports Medicine | 438 | 378 | 86.301 | 9,565 | 25.304 |
| 6 | Annals of Botany | 209 | 199 | 95.215 | 2,866 | 14.402 |
| 7 | Applied Catalysis B Environmental | 788 | 305 | 38.706 | 741 | 2.43 |
| 8 | Bioinformatics | 811 | 808 | 99.63 | 9,454 | 11.7 |
| 9 | Biomaterials | 651 | 326 | 50.077 | 831 | 2.549 |
| 10 | Biophysical Journal | 3,644 | 468 | 12.843 | 2,347 | 5.015 |
| 11 | Biosensors & Bioelectronics | 1,014 | 635 | 62.623 | 1,471 | 2.317 |
| 12 | Briefings in Bioinformatics | 93 | 93 | 100 | 1,159 | 12.462 |
| 13 | Cell Reports | 1,100 | 1,071 | 97.364 | 16,028 | 14.965 |
| 14 | Cell | 661 | 642 | 97.126 | 37,557 | 58.5 |
| 15 | Chinese Journal of Catalysis | 239 | 2 | 0.837 | 2 | 1 |
| 16 | Clinical Infectious Diseases | 777 | 535 | 68.855 | 10,686 | 19.974 |
| 17 | Current Biology | 846 | 796 | 94.09 | 31,275 | 39.29 |
| 18 | Database The Journal of Biological | 164 | 123 | 75 | 887 | 7.211 |
|    | Databases And Curation                               |        |        |        |         |         |
| 19 | European Journal of Sport Science                    | 155    | 153    | 98.71  | 4,114   | 26.889  |
| 20 | Expert Review of Medical Devices                     | 116    | 116    | 100    | 203     | 1.75    |
| 21 | International Journal of Production Research         | 460    | 23     | 5      | 48      | 2.087   |
| 22 | International Journal of Remote Sensing              | 334    | 30     | 8.982  | 206     | 6.867   |
| 23 | Journal of Catalysis                                 | 358    | 38     | 10.615 | 93      | 2.447   |
| 24 | Journal of Experimental Botany                       | 514    | 467    | 90.856 | 2,476   | 5.302   |
| 25 | Journal of Infectious Diseases                       | 691    | 515    | 74.53  | 4,398   | 8.54    |
| 26 | Journal of Power Sources                             | 1,651  | 93     | 5.633  | 221     | 2.376   |
| 27 | Journal of Sports Sciences                           | 297    | 290    | 97.643 | 6,977   | 24.059  |
| 28 | Journal of The American Statistical Association      | 161    | 158    | 98.137 | 605     | 3.829   |
| 29 | Materials Today                                      | 96     | 21     | 21.875 | 109     | 5.19    |
| 30 | Molecular Ecology                                    | 433    | 415    | 95.843 | 5,541   | 13.352  |
| 31 | Nano Letters                                         | 1,194  | 775    | 64.908 | 6,287   | 8.112   |
| 32 | Nanotoxicology                                       | 150    | 62     | 41.333 | 138     | 2.226   |
| 33 | Nature Communications                                | 3,685  | 3,132  | 84.993 | 76,607  | 24.459  |
| 34 | Nature Materials                                     | 277    | 222    | 80.144 | 3,199   | 14.41   |
| 35 | Nature                                               | 2,397  | 2,386  | 99.541 | 487,961 | 204.51  |
| 36 | New Phytologist                                      | 530    | 522    | 98.491 | 7,172   | 13.739  |
| 37 | NPG Asia Materials                                   | 104    | 98     | 94.231 | 242     | 2.469   |
| 38 | Nucleic Acids Research                               | 1,298  | 1,144  | 88.136 | 8,157   | 7.13    |
| 39 | Plant Biotechnology Journal                          | 203    | 148    | 72.906 | 912     | 6.162   |
| 40 | Plant, Cell And Environment                          | 220    | 78     | 35.455 | 325     | 4.167   |
| 41 | PLoS Computational Biology                           | 582    | 576    | 98.969 | 14,629  | 25.398  |
| 42 | PLoS Neglected Tropical Diseases                     | 876    | 843    | 96.233 | 10,954  | 12.994  |
| 43 | PLoS One                                             | 23,019 | 16,942 | 73.6   | 160,665 | 9.483   |
| 44 | PLoS Pathogens                                       | 684    | 676    | 98.83  | 9,968   | 14.746  |
| 45 | Progress in Materials Science                        | 49     | 18     | 36.735 | 86      | 4.778   |
| 46 | Progress in Photovoltaics                            | 143    | 18     | 12.587 | 37      | 2.056   |
| 47 | Progress in Polymer Science                          | 49     | 10     | 20.408 | 29      | 2.9     |
| 48 | Scientific Reports                                   | 21,057 | 12,442 | 59.087 | 124,815 | 10.032  |
| 49 | Small                                                | 659    | 187    | 28.376 | 378     | 2.021   |
| 50 | Statistical Methods in Medical Research              | 189    | 100    | 52.91  | 309     | 3.09    |
|    |                                                      |        | Mean   | 62.3   |         | 14.135  |
|    |                                                      |        | Median | 73.253 |         | 6.515   |

**Table 4: Mention statistics for journals without Twitter plugin**

| S. No. | Journal Name | Records with DOI | Tweeted | Coverage % | Twitter Mentions | Avg. Twitter Mentions |
|---|---|---|---|---|---|---|
| 1 | Acta Mathematica Sinica | 116 | 6 | 5.172 | 31 | 5.167 |
| 2 | Advances in Mathematical Physics | 122 | 15 | 12.295 | 40 | 2.667 |
| 3 | Aircraft Engineering and Aerospace Technology | 90 | 1 | 1.111 | 2 | 2 |
| 4 | Analytical & Bioanalytical Chemistry | 837 | 486 | 58.065 | 1,116 | 2.296 |
| 5 | Annals of Biomedical Engineering | 305 | 118 | 38.689 | 310 | 2.627 |
| 6 | Annals of Operations Research | 312 | 25 | 8.013 | 64 | 2.56 |
| 7 | Applied Microbiology & Biotechnology | 887 | 433 | 48.816 | 970 | 2.24 |
| 8 | Autonomous Robots | 78 | 78 | 100 | 183 | 2.346 |
| 9 | Biochemistry-Moscow | 171 | 31 | 18.129 | 73 | 2.355 |
| 10 | Biomed Research International | 1,852 | 518 | 27.97 | 1,378 | 2.66 |
| 11 | British Food Journal | 188 | 24 | 12.766 | 81 | 3.375 |
| 12 | Bulletin of Mathematical Biology | 95 | 45 | 47.368 | 163 | 3.622 |
| 13 | Canadian Respiratory Journal | 95 | 34 | 35.789 | 72 | 2.118 |
| 14 | Chinese Annals of Mathematics | 73 | 0 | 0 | 0 | |
| 15 | Complexity | 194 | 45 | 23.196 | 425 | 9.444 |
| 16 | Disease Markers | 161 | 45 | 27.95 | 73 | 1.622 |
| 17 | Empirical Software Engineering | 73 | 22 | 30.137 | 132 | 6 |
| 18 | IEEE Access | 815 | 9 | 1.104 | 30 | 3.333 |
| 19 | IEEE Communications Letters | 635 | 43 | 6.772 | 77 | 1.791 |
| 20 | IEEE Journal on Selected Areas in Communications | 301 | 33 | 10.963 | 72 | 2.182 |
| 21 | IEEE Transactions on Antennas and Propagation | 677 | 35 | 5.17 | 52 | 1.486 |
| 22 | IEEE Transactions on Biomedical Engineering | 269 | 75 | 27.881 | 271 | 3.613 |
| 23 | IEEE Transactions on Communications | 405 | 54 | 13.333 | 113 | 2.093 |
| 24 | IEEE Transactions on Geoscience and Remote Sensing | 573 | 60 | 10.471 | 115 | 1.917 |
| 25 | IEEE Transactions on Image Processing | 446 | 82 | 18.386 | 272 | 3.317 |
| 26 | IEEE Transactions on Industrial Electronics | 765 | 15 | 1.961 | 19 | 1.267 |
| 27 | IEEE Transactions on Mobile Computing | 224 | 10 | 4.464 | 43 | 4.3 |
| 28 | IEEE Transactions on Neural Networks and Learning Systems | 228 | 46 | 20.175 | 96 | 2.087 |
| 29 | IEEE Transactions on Pattern Analysis and Machine Intelligence | 193 | 119 | 61.658 | 298 | 2.504 |
| 30 | IEEE Transactions on Power Electronics | 736 | 18 | 2.446 | 36 | 2 |
| 31 | IEEE Transactions on Power Systems | 528 | 39 | 7.386 | 120 | 3.077 |
| 32 | IEEE Transactions on Signal Processing | 486 | 184 | 37.86 | 567 | 3.082 |
| 33 | IEEE Transactions on Visualization and Computer Graphics | 240 | 46 | 19.167 | 136 | 2.957 |
| 34 | IEEE Transactions on Wireless Communications | 631 | 94 | 14.897 | 193 | 2.053 |
| 35 | Industrial Management & Data Systems | 96 | 10 | 10.417 | 22 | 2.2 |
| 36 | International Journal of Advanced Manufacturing Technology | 1,420 | 22 | 1.549 | 33 | 1.5 |
| 37 | International Journal of Numerical Methods for Heat & Fluid Flow | 132 | 1 | 0.758 | 1 | 1 |
| 38 | Journal of Lightwave Technology | 691 | 59 | 8.538 | 142 | 2.407 |
| 39 | Journal of Mathematical Biology | 133 | 50 | 37.594 | 117 | 2.34 |
| 40 | Kybernetes | 101 | 5 | 4.95 | 15 | 3 |
| 41 | Multimedia Tools & | 876 | 41 | 4.68 | 84 | 2.049 |

|    | Applications | | | | | |
|----|---|---|---|---|---|---|
| 42 | Online Information Review | 68 | 25 | 36.765 | 123 | 4.92 |
| 43 | Parasitology Research | 532 | 429 | 80.639 | 1092 | 2.545 |
| 44 | Proceedings of The IEEE | 155 | 39 | 25.161 | 421 | 10.795 |
| 45 | Rapid Prototyping Journal | 90 | 4 | 4.444 | 7 | 1.75 |
| 46 | Scanning | 102 | 15 | 14.706 | 34 | 2.267 |
| 47 | Security and Communication Networks | 475 | 26 | 5.474 | 34 | 1.308 |
| 48 | Shock and Vibration | 390 | 5 | 1.282 | 6 | 1.2 |
| 49 | Springerplus | 2,115 | 601 | 28.416 | 2345 | 3.902 |
| 50 | Surgical Endoscopy & Other Interventional Techniques | 752 | 457 | 60.771 | 1823 | 3.989 |
| | | | **Mean** | **21.714** | | **2.925** |
| | | | **Median** | **14.02** | | **2.355** |

**Table 5: Mention statistics for journals with Facebook plugin**

| S. No. | Journal Name | Records with DOI | Found in FB | Coverage % | FB Mentions | Avg. FB Mentions |
|---|---|---|---|---|---|---|
| 1 | Acs Nano | 1,285 | 104 | 8.093 | 168 | 1.615 |
| 2 | Advanced Energy Materials | 419 | 5 | 1.193 | 5 | 1 |
| 3 | Advanced Functional Materials | 881 | 19 | 2.157 | 23 | 1.211 |
| 4 | Advanced Materials | 1,169 | 7 | 0.599 | 10 | 1.429 |
| 5 | American Journal of Sports Medicine | 438 | 216 | 49.315 | 929 | 4.301 |
| 6 | Annals of Botany | 209 | 126 | 60.287 | 255 | 2.024 |
| 7 | Applied Catalysis B Environmental | 788 | 9 | 1.142 | 11 | 1.222 |
| 8 | Bioinformatics | 811 | 153 | 18.866 | 183 | 1.196 |
| 9 | Biomaterials | 651 | 56 | 8.602 | 78 | 1.393 |
| 10 | Biophysical Journal | 3,644 | 36 | 0.988 | 49 | 1.361 |
| 11 | Biosensors & Bioelectronics | 1,014 | 22 | 2.17 | 25 | 1.136 |
| 12 | Briefings in Bioinformatics | 93 | 14 | 15.054 | 18 | 1.286 |
| 13 | Cell Reports | 1,100 | 424 | 38.545 | 939 | 2.215 |
| 14 | Cell | 661 | 403 | 60.968 | 1,751 | 4.345 |
| 15 | Chinese Journal of Catalysis | 239 | 0 | 0 | 0 | 0 |
| 16 | Clinical Infectious Diseases | 777 | 185 | 23.81 | 495 | 2.676 |
| 17 | Current Biology | 846 | 325 | 38.416 | 1,316 | 4.049 |
| 18 | Database The Journal of Biological Databases And Curation | 164 | 20 | 12.195 | 26 | 1.3 |
| 19 | European Journal of Sport Science | 155 | 73 | 47.097 | 290 | 3.973 |
| 20 | Expert Review of Medical Devices | 116 | 7 | 6.034 | 8 | 1.143 |
| 21 | International Journal of Production Research | 460 | 3 | 0.652 | 3 | 1 |
| 22 | International Journal of Remote Sensing | 334 | 17 | 5.09 | 20 | 1.176 |
| 23 | Journal of Catalysis | 358 | 4 | 1.117 | 4 | 1 |
| 24 | Journal of Experimental Botany | 514 | 93 | 18.093 | 139 | 1.495 |
| 25 | Journal of Infectious Diseases | 691 | 136 | 19.682 | 255 | 1.875 |
| 26 | Journal of Power Sources | 1,651 | 18 | 1.09 | 21 | 1.167 |
| 27 | Journal of Sports Sciences | 297 | 89 | 29.966 | 185 | 2.079 |
| 28 | Journal of The American Statistical Association | 161 | 2 | 1.242 | 2 | 1 |
| 29 | Materials Today | 96 | 11 | 11.458 | 15 | 1.364 |
| 30 | Molecular Ecology | 433 | 96 | 22.171 | 204 | 2.125 |
| 31 | Nano Letters | 1,194 | 74 | 6.198 | 149 | 2.014 |
| 32 | Nanotoxicology | 150 | 2 | 1.333 | 3 | 1.5 |
| 33 | Nature Communications | 3,685 | 1,068 | 28.982 | 3,356 | 3.142 |
| 34 | Nature Materials | 277 | 65 | 23.466 | 189 | 2.908 |
| 35 | Nature | 2,397 | 1,936 | 80.768 | 19,927 | 10.293 |
| 36 | New Phytologist | 530 | 218 | 41.132 | 452 | 2.073 |
| 37 | NPG Asia Materials | 104 | 12 | 11.538 | 18 | 1.5 |
| 38 | Nucleic Acids Research | 1,298 | 134 | 10.324 | 186 | 1.388 |
| 39 | Plant Biotechnology Journal | 203 | 65 | 32.02 | 129 | 1.985 |
| 40 | Plant, Cell And Environment | 220 | 16 | 7.273 | 21 | 1.312 |
| 41 | PLoS Computational Biology | 582 | 164 | 28.179 | 324 | 1.976 |
| 42 | PLoS Neglected Tropical Diseases | 876 | 340 | 38.813 | 700 | 2.059 |
| 43 | PLoS One | 23,019 | 3,901 | 16.947 | 9,231 | 2.366 |
| 44 | PLoS Pathogens | 684 | 210 | 30.702 | 418 | 1.99 |
| 45 | Progress in Materials Science | 49 | 1 | 2.041 | 1 | 1 |
| 46 | Progress in Photovoltaics | 143 | 6 | 4.196 | 6 | 1 |
| 47 | Progress in Polymer Science | 49 | 5 | 10.204 | 5 | 1 |
| 48 | Scientific Reports | 21,057 | 2,540 | 12.062 | 6,596 | 2.597 |

| 49 | Small | 659 | 25 | 3.794 | 29 | 1.16 |
| 50 | Statistical Methods in Medical Research | 189 | 2 | 1.058 | 2 | 1 |
| | | | **Mean** | **17.942** | | **1.948** |
| | | | **Median** | **11.498** | | **1.462** |

**Table 6: Mention statistics for journals without Facebook plugin**

| S. No. | Journal Name | Records with DOI | Found in FB | Coverage % | FB Mentions | Avg. FB Mentions |
|---|---|---|---|---|---|---|
| 1 | Acta Mathematica Sinica | 116 | 0 | 0 | 0 | 0 |
| 2 | Advances in Mathematical Physics | 122 | 2 | 1.639 | 2 | 1 |
| 3 | Aircraft Engineering and Aerospace Technology | 90 | 0 | 0 | 0 | 0 |
| 4 | Analytical & Bioanalytical Chemistry | 837 | 26 | 3.106 | 38 | 1.462 |
| 5 | Annals of Biomedical Engineering | 305 | 17 | 5.574 | 26 | 1.529 |
| 6 | Annals of Operations Research | 312 | 2 | 0.641 | 2 | 1 |
| 7 | Applied Microbiology & Biotechnology | 887 | 26 | 2.931 | 48 | 1.846 |
| 8 | Autonomous Robots | 78 | 65 | 83.333 | 65 | 1 |
| 9 | Biochemistry-Moscow | 171 | 3 | 1.754 | 3 | 1 |
| 10 | Biomed Research International | 1,852 | 121 | 6.533 | 215 | 1.777 |
| 11 | British Food Journal | 188 | 9 | 4.787 | 11 | 1.222 |
| 12 | Bulletin of Mathematical Biology | 95 | 15 | 15.789 | 18 | 1.2 |
| 13 | Canadian Respiratory Journal | 95 | 6 | 6.316 | 6 | 1 |
| 14 | Chinese Annals of Mathematics | 73 | 0 | 0 | 0 | 0 |
| 15 | Complexity | 194 | 17 | 8.763 | 33 | 1.941 |
| 16 | Disease Markers | 161 | 7 | 4.348 | 7 | 1 |
| 17 | Empirical Software Engineering | 73 | 67 | 91.781 | 82 | 1.224 |
| 18 | IEEE Access | 815 | 1 | 0.123 | 1 | 1 |
| 19 | IEEE Communications Letters | 635 | 0 | 0 | 0 | 0 |
| 20 | IEEE Journal on Selected Areas in Communications | 301 | 0 | 0 | 0 | 0 |
| 21 | IEEE Transactions on Antennas and Propagation | 677 | 5 | 0.739 | 6 | 1.2 |
| 22 | IEEE Transactions on Biomedical Engineering | 269 | 9 | 3.346 | 10 | 1.111 |
| 23 | IEEE Transactions on Communications | 405 | 0 | 0 | 0 | 0 |
| 24 | IEEE Transactions on Geoscience and | 573 | 5 | 0.873 | 6 | 1.2 |
| | Remote Sensing | | | | | |
| 25 | IEEE Transactions on Image Processing | 446 | 1 | 0.224 | 2 | 2 |
| 26 | IEEE Transactions on Industrial Electronics | 765 | 0 | 0 | 0 | 0 |
| 27 | IEEE Transactions on Mobile Computing | 224 | 2 | 0.893 | 2 | 1 |
| 28 | IEEE Transactions on Neural Networks and Learning Systems | 228 | 7 | 3.07 | 8 | 1.143 |
| 29 | IEEE Transactions on Pattern Analysis and Machine Intelligence | 193 | 3 | 1.554 | 5 | 1.667 |
| 30 | IEEE Transactions on Power Electronics | 736 | 1 | 0.136 | 1 | 1 |
| 31 | IEEE Transactions on Power Systems | 528 | 0 | 0 | 0 | 0 |
| 32 | IEEE Transactions on Signal Processing | 486 | 4 | 0.823 | 4 | 1 |
| 33 | IEEE Transactions on Visualization and Computer Graphics | 240 | 3 | 1.25 | 4 | 1.333 |
| 34 | IEEE Transactions on Wireless Communications | 631 | 0 | 0 | 0 | 0 |
| 35 | Industrial Management & Data Systems | 96 | 4 | 4.167 | 13 | 3.25 |
| 36 | International Journal of Advanced Manufacturing Technology | 1,420 | 7 | 0.493 | 7 | 1 |
| 37 | International Journal of Numerical Methods for Heat & Fluid Flow | 132 | 0 | 0 | 0 | 0 |
| 38 | Journal of Lightwave Technology | 691 | 0 | 0 | 0 | 0 |
| 39 | Journal of Mathematical Biology | 133 | 4 | 3.008 | 6 | 1.5 |
| 40 | Kybernetes | 101 | 0 | 0 | 0 | 0 |
| 41 | Multimedia Tools & Applications | 876 | 7 | 0.799 | 8 | 1.143 |
| 42 | Online Information Review | 68 | 4 | 5.882 | 14 | 3.5 |
| 43 | Parasitology Research | 532 | 36 | 6.767 | 56 | 1.556 |
| 44 | Proceedings of The IEEE | 155 | 58 | 37.419 | 91 | 1.569 |
| 45 | Rapid Prototyping Journal | 90 | 0 | 0 | 0 | 0 |
| 46 | Scanning | 102 | 0 | 0 | 0 | 0 |
| 47 | Security and Communication Networks | 475 | 5 | 1.053 | 9 | 1.8 |
| 48 | Shock and Vibration | 390 | 0 | 0 | 0 | 0 |
| 49 | Springerplus | 2,115 | 137 | 6.478 | 244 | 1.781 |
| 50 | Surgical Endoscopy & Other Interventional Techniques | 752 | 364 | 48.404 | 391 | 1.074 |
| | | | **Mean** | 7.296 | | 1.001 |
| | | | **Median** | 0.973 | | 1 |

**Table 7: Mention statistics for journals with Mendeley plugin**

| S. No. | Journal Name | Records with DOI | Found in Mendeley | Coverage % | Mendeley Reads | Avg. Mendeley Reads |
|---|---|---|---|---|---|---|
| 1 | Acs Nano | 1,285 | 975 | 75.875 | 61,788 | 63.372 |
| 2 | American Journal of Sports Medicine | 438 | 382 | 87.215 | 20,814 | 54.487 |
| 3 | Annals of Botany | 209 | 208 | 99.522 | 8,478 | 40.76 |
| 4 | Bioinformatics | 811 | 809 | 99.753 | 43,247 | 53.457 |
| 5 | Briefings in Bioinformatics | 93 | 93 | 100 | 9,005 | 96.828 |
| 6 | Clinical Infectious Diseases | 777 | 624 | 80.309 | 28,287 | 45.332 |
| 7 | Database The Journal of Biological Databases And Curation | 164 | 136 | 82.927 | 5,142 | 37.809 |
| 8 | European Journal of Sport Science | 155 | 153 | 98.71 | 8,048 | 52.601 |
| 9 | Expert Review of Medical Devices | 116 | 113 | 97.414 | 2,589 | 22.912 |
| 10 | International Journal of Production Research | 460 | 116 | 25.217 | 5,999 | 51.716 |
| 11 | International Journal of Remote Sensing | 334 | 84 | 25.15 | 2,037 | 24.25 |
| 12 | Journal of Experimental Botany | 514 | 485 | 94.358 | 24,359 | 50.225 |
| 13 | Journal of Infectious Diseases | 691 | 609 | 88.133 | 22,029 | 36.172 |
| 14 | Journal of Sports Sciences | 297 | 291 | 97.98 | 18,555 | 63.763 |
| 15 | Journal of The American Statistical Association | 161 | 153 | 95.031 | 3,726 | 24.353 |
| 16 | Nano Letters | 1,194 | 991 | 82.998 | 63,861 | 64.441 |
| 17 | Nanotoxicology | 150 | 93 | 62 | 3,020 | 32.473 |
| 18 | Nucleic Acids Research | 1,298 | 1,222 | 94.145 | 91,737 | 75.071 |
| 19 | PLoS Computational Biology | 582 | 580 | 99.656 | 38,243 | 65.936 |
| 20 | PLoS Neglected Tropical Diseases | 876 | 855 | 97.603 | 47,230 | 55.24 |
| 21 | PLoS One | 23,019 | 19,462 | 84.548 | 719,920 | 36.991 |
| 22 | PLoS Pathogens | 684 | 683 | 99.854 | 30,213 | 44.236 |
| | | | **Mean** | **84.927** | | **49.656** |
| | | | **Median** | **94.252** | | **50.971** |

Table 8: Mention statistics for journals without Mendeley plugin

| S. No. | Journal Name | Records with DOI | Found in Mendeley | Coverage % | Mendeley Reads | Avg. Mendeley Reads |
|---|---|---|---|---|---|---|
| 1 | Acta Mathematica Sinica | 116 | 6 | 5.172 | 39 | 6.5 |
| 2 | Advances in Mathematical Physics | 122 | 17 | 13.934 | 73 | 4.294 |
| 3 | Aircraft Engineering and Aerospace Technology | 90 | 5 | 5.556 | 39 | 7.8 |
| 4 | Analytical & Bioanalytical Chemistry | 837 | 533 | 63.68 | 11,270 | 21.144 |
| 5 | Annals of Biomedical Engineering | 305 | 198 | 64.918 | 7,477 | 37.763 |
| 6 | Annals of Operations Research | 312 | 67 | 21.474 | 2,364 | 35.284 |
| 7 | Applied Microbiology & Biotechnology | 887 | 519 | 58.512 | 19,765 | 38.083 |
| 8 | Autonomous Robots | 78 | 78 | 100 | 2,802 | 35.923 |
| 9 | Biochemistry-Moscow | 171 | 45 | 26.316 | 874 | 19.422 |
| 10 | Biomed Research International | 1,852 | 753 | 40.659 | 24,179 | 32.11 |
| 11 | British Food Journal | 188 | 60 | 31.915 | 2,315 | 38.583 |
| 12 | Bulletin of Mathematical Biology | 95 | 61 | 64.211 | 1,119 | 18.344 |
| 13 | Canadian Respiratory Journal | 95 | 48 | 50.526 | 1,110 | 23.125 |
| 14 | Chinese Annals of Mathematics | 73 | 1 | 1.37 | 2 | 2 |
| 15 | Complexity | 194 | 56 | 28.866 | 646 | 11.536 |
| 16 | Disease Markers | 161 | 67 | 41.615 | 1,520 | 22.687 |
| 17 | Empirical Software Engineering | 73 | 68 | 93.151 | 2,993 | 44.015 |
| 18 | IEEE Access | 815 | 23 | 2.822 | 5,680 | 246.957 |
| 19 | IEEE Communications Letters | 635 | 126 | 19.843 | 1,827 | 14.5 |
| 20 | IEEE Journal on Selected Areas in Communications | 301 | 109 | 36.213 | 4,445 | 40.78 |
| 21 | IEEE Transactions on Antennas and Propagation | 677 | 147 | 21.713 | 2,055 | 13.98 |
| 22 | IEEE Transactions on Biomedical Engineering | 269 | 156 | 57.993 | 7,870 | 50.449 |
| 23 | IEEE Transactions on Communications | 405 | 141 | 34.815 | 3,456 | 24.511 |
| 24 | IEEE Transactions on Geoscience and Remote Sensing | 573 | 146 | 25.48 | 3,758 | 25.74 |
| 25 | IEEE Transactions on Image Processing | 446 | 183 | 41.031 | 6,182 | 33.781 |
| 26 | IEEE Transactions on Industrial Electronics | 765 | 154 | 20.131 | 4,636 | 30.104 |
| 27 | IEEE Transactions on Mobile Computing | 224 | 93 | 41.518 | 2,334 | 25.097 |
| 28 | IEEE Transactions on Neural Networks and Learning Systems | 228 | 92 | 40.351 | 3,205 | 34.837 |
| 29 | IEEE Transactions on Pattern Analysis and Machine Intelligence | 193 | 142 | 73.575 | 15,652 | 110.225 |
| 30 | IEEE Transactions on Power Electronics | 736 | 143 | 19.429 | 5,599 | 39.154 |
| 31 | IEEE Transactions on Power Systems | 528 | 149 | 28.22 | 6,198 | 41.597 |

| # | Journal | | | | |
|---|---|---|---|---|---|
| 32 | IEEE Transactions on Signal Processing | 486 | 249 | 51.235 | 6,386 | 25.647 |
| 33 | IEEE Transactions on Visualization and Computer Graphics | 240 | 82 | 34.167 | 6,192 | 75.512 |
| 34 | IEEE Transactions on Wireless Communications | 631 | 207 | 32.805 | 5,641 | 27.251 |
| 35 | Industrial Management & Data Systems | 96 | 30 | 31.25 | 4,452 | 148.4 |
| 36 | International Journal of Advanced Manufacturing Technology | 1,420 | 312 | 21.972 | 7,649 | 24.516 |
| 37 | International Journal of Numerical Methods for Heat & Fluid Flow | 132 | 11 | 8.333 | 67 | 6.091 |
| 38 | Journal of Lightwave Technology | 691 | 155 | 22.431 | 3,593 | 23.181 |
| 39 | Journal of Mathematical Biology | 133 | 70 | 52.632 | 1,145 | 16.357 |
| 40 | Kybernetes | 101 | 7 | 6.931 | 137 | 19.571 |
| 41 | Multimedia Tools & Applications | 876 | 102 | 11.644 | 1,890 | 18.529 |
| 42 | Online Information Review | 68 | 28 | 41.176 | 2,031 | 72.536 |
| 43 | Parasitology Research | 532 | 439 | 82.519 | 8,787 | 20.016 |
| 44 | Proceedings of The IEEE | 155 | 92 | 59.355 | 9,812 | 106.652 |
| 45 | Rapid Prototyping Journal | 90 | 9 | 10 | 471 | 52.333 |
| 46 | Scanning | 102 | 21 | 20.588 | 338 | 16.095 |
| 47 | Security and Communication Networks | 475 | 66 | 13.895 | 794 | 12.03 |
| 48 | Shock and Vibration | 390 | 44 | 11.282 | 387 | 8.795 |
| 49 | Springerplus | 2,115 | 833 | 39.385 | 20,575 | 24.7 |
| 50 | Surgical Endoscopy & Other Interventional Techniques | 752 | 648 | 86.17 | 13,970 | 21.559 |
| 51 | Advanced Energy Materials | 419 | 93 | 22.196 | 7,159 | 76.978 |
| 52 | Advanced Functional Materials | 881 | 317 | 35.982 | 16,537 | 52.167 |
| 53 | Advanced Materials | 1,169 | 69 | 5.902 | 4,934 | 71.507 |
| 54 | Applied Catalysis B Environmental | 788 | 347 | 44.036 | 14,341 | 41.329 |
| 55 | Biomaterials | 651 | 449 | 68.971 | 26,121 | 58.176 |
| 56 | Biophysical Journal | 3,644 | 633 | 17.371 | 17,149 | 27.092 |
| 57 | Biosensors & Bioelectronics | 1,014 | 697 | 68.738 | 29,513 | 42.343 |
| 58 | Cell Reports | 1,100 | 1,079 | 98.091 | 78,027 | 72.314 |
| 59 | Cell | 661 | 647 | 97.882 | 168,245 | 260.039 |
| 60 | Chinese Journal of Catalysis | 239 | 8 | 3.347 | 204 | 25.5 |
| 61 | Current Biology | 846 | 821 | 97.045 | 62,638 | 76.295 |
| 62 | Journal of Catalysis | 358 | 87 | 24.302 | 3,852 | 44.276 |
| 63 | Journal of Power Sources | 1,651 | 315 | 19.079 | 14,383 | 45.66 |
| 64 | Materials Today | 96 | 65 | 67.708 | 7,344 | 112.985 |
| 65 | Molecular Ecology | 433 | 423 | 97.691 | 40,665 | 96.135 |
| 66 | Nature Communications | 3,685 | 3,431 | 93.107 | 300,482 | 87.579 |

| 67 | Nature Materials | 277 | 241 | 87.004 | 39,042 | 162 |
| 68 | Nature | 2,397 | 2,329 | 97.163 | 339,385 | 145.721 |
| 69 | New Phytologist | 530 | 523 | 98.679 | 35,671 | 68.205 |
| 70 | NPG Asia Materials | 104 | 98 | 94.231 | 3,446 | 35.163 |
| 71 | Plant Biotechnology Journal | 203 | 173 | 85.222 | 10,053 | 58.11 |
| 72 | Plant, Cell And Environment | 220 | 97 | 44.091 | 3,841 | 39.598 |
| 73 | Progress in Materials Science | 49 | 24 | 48.98 | 3,714 | 154.75 |
| 74 | Progress in Photovoltaics | 143 | 45 | 31.469 | 1,012 | 22.489 |
| 75 | Progress in Polymer Science | 49 | 22 | 44.898 | 3,287 | 149.409 |
| 76 | Scientific Reports | 21,057 | 14,846 | 70.504 | 604,874 | 40.743 |
| 77 | Small | 659 | 296 | 44.917 | 10,276 | 34.716 |
| 78 | Statistical Methods in Medical Research | 189 | 130 | 68.783 | 3,437 | 26.438 |
| | | | **Mean** | **36.256** | | **37.002** |
| | | | **Median** | **32.36** | | **24.899** |